\documentclass[sn-basic]{sn-jnl}

\jyear{2021}%

\begin{document}

\title[Shape descriptors for nodal forces]{Spherical harmonic shape descriptors of nodal force demands for quantifying spatial truss connection complexity}

\author*[1]{\fnm{Keith J.} \sur{Lee}}\email{keithjl@mit.edu}
\author[1]{\fnm{Renaud} \sur {Danhaive}}\email{danhaive@mit.edu}
\author[1]{\fnm{Caitlin T.} \sur{Mueller}}\email{caitlinm@mit.edu}

\affil*[1]{\orgdiv{Department of Architecture, Building Technology}, \orgname{Massachusetts Institute of Technology}, \orgaddress{\street{77 Massachusetts Avenue}, \city{Cambridge}, \postcode{02139}, \state{Massachusetts}, \country{USA}}}

\abstract{The connections of a spatial truss structure play a critical role in the safe and efficient transfer of axial forces between members. For discrete connections, they can also improve construction efficiency by acting as registration devices that lock members in precise orientations. As more geometrically complex spatial trusses are enabled by computational workflows and the demand for material-efficient spanning systems, there is a need to understand the effects of global form on the demands at the connections. For large-scale structures with irregular geometry, customizing individual nodes to meet exact member orientations and force demands may be infeasible; conversely, standardizing all connections results in oversized nodes and a compromise in registration potential. We propose a method for quantifying the complexity of spatial truss designs by the variation in nodal force demands. By representing nodal forces as a geometric object, we leverage the spherical harmonic shape descriptor, developed for applications in computational geometry, to characterize each node by a rotation and translation-invariant fixed-length vector. We define a complexity score for spatial truss design by the variance in the positions of the feature vectors in higher-dimensional space, providing an additional performance metric during early stage design exploration. We then develop a pathway towards reducing complexity by clustering nodes with respect to their feature vectors to reduce the number of unique connectors for design while minimizing the effects of mass standardization.}


\maketitle

\section{Introduction}\label{sec:1}
\begin{figure}[h]
    \centering
    \includegraphics[width = \textwidth]{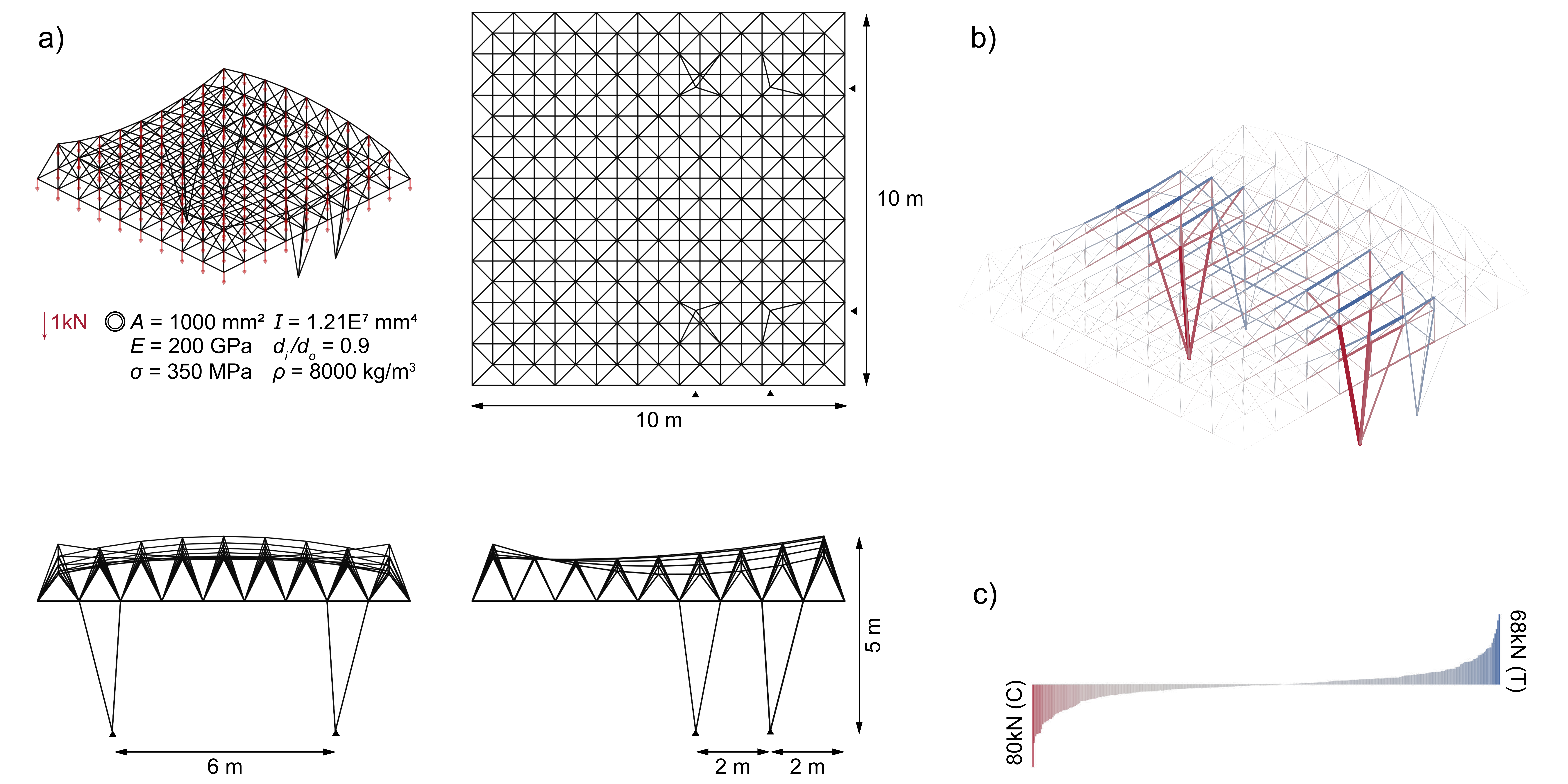}
    \caption{A spatial truss design: a) problem formulation, b) displaced shape with internal force colormap, c) distribution of axial forces}
    \label{fig:problem}
\end{figure}

The complexity of a structural design is often the restricting factor to its realization. It is also difficult to define and quantify. Variations in internal forces to be resisted and the optimal member sizes required, anticipated fabrication time and tooling, and the ease of assembly all contribute to design complexity. For spatial truss structures, all of these measures apply. Consider the spatial truss in Figure \ref{fig:problem}, consisting of 185 nodes and 664 tubular steel elements. Despite a regular planar grid spacing of nodes and relatively regular member orientations, a significant variation in axial forces is observed.

For many similar structures built around the world, two missed opportunities are often evident: first, members are typically uniformly sized, resulting in unnecessary material consumption and the increase in structural mass. Often, this uniform sizing comes from the reduction of material procurement complexity and resulting cost reduction. Second, the design and detailing of nodal connections are also uniform, despite an equally large variation in force transfer requirements. Generally, these nodes are often heavy steel cylinders or spheres with mitered element ends and welded connections. Although the connections themselves come from easily fabricated parts, construction complexity is increased by the lack of inherent registration capacity of the nodes, as well as the time required to miter weld the elements. Further, as the building industry pushes towards facilitating the deconstruction and reuse of building components, providing discrete nodes over welded connections should be encouraged \citep{fivet2020nothing, brutting2019design}.

Although the optimal sizing and orientation of spatial truss elements has been the focus of much research, the performance potential of the connections has been less understood. Spatial truss nodes play an important role, both as member registration devices that lock member orientations and facilitate assembly, and as a critical force transfer mechanism between elements. For shape-optimized trusses, where orientations are often highly irregular and transferred forces are large, the need for both registration and strength is critical. 

Ideally, all nodes in a spatial truss should be tailored to their connected members and anticipated forces. However, this adds significant complexity to all stages of the design, manufacture, and construction process. Conversely, if a single standardized connection is used, it must be oversized for most of the structure, and providing registration for elements would be difficult for non-regular truss geometries. In this case, complexity is inherent in the variation of the force demands at each node, and represents the lost potential of customization. If the variations in nodal demands can be quantified, designers can strategically reduce the anticipated complexity of design and manufacture by identifying critical nodes that warrant customization, and nodes where a standardized connector is sufficient.

We develop a method of consistent nodal demand characterization and its resulting nodal complexity metric to provide this insight to designers. By representing the forces acting at a node by a single spherical object, we leverage the spherical harmonic shape descriptor developed in the field of computational geometry to characterize an arbitrary number of force magnitudes and orientations with a fixed-length feature vector. The conversion of forces into a single geometric object resolves challenges in comparing the similarity of nodes with different numbers of element connectivity, and the use of the spherical harmonic shape descriptor provides invariance to the orientation of the node, allowing for recognition of similar nodal demands that are rotated or translated throughout a spatial truss. The feature vectors of all nodes provide the basis of a nodal distance matrix, which is then used for a complexity metric when evaluating multiple designs, and for clustering analysis to strategically reduce the number of unique connections required for design and fabrication. 

\section{Related work}\label{sec:2}
Our method builds upon existing research in quantifying design complexity, design rationalization, computational geometry, and geometric representation of forces. We present several key works in each field.

\subsection{Design complexity and rationalization}\label{subsec:2_1}
In mechanical and industrial design, \cite{bashir_estimating_1999} defines a quantitative metric of design complexity by the depth and breadth of the function trees of each component in the product; the authors find a positive correlation with worker-hours required at surveyed manufacturing plants. Design for Assembly and Assembly-Oriented Design procedures have been developed to integrate assembly feasibility in the early design phase of product design \citep{boothroyd_design_1987,dalgleish_design_2000}. These procedures range from qualitative rubrics for engineering feedback to rigorous data management for communication between engineers, designers, and fabricators \citep{demoly2011assembly}.

In the design of buildings, \cite{mitchell2005constructing} qualitatively defines architectural complexity as the ratio of the quantity of information required to geometrically represent the design to the quantity of information required for fabrication and assembly of the elements. These two measures are presented as conceptually orthogonal axes, where both components can either be minimized (highly regular, modular construction) or maximized (highly customized, complex construction). \cite{scheurer2010materialising} provides insight on the relationship between geometric complexity and the technology required for its realization from the perspective of an architecture practitioner.

Reducing complexity in the manufacturing and construction process has a significant impact on material consumption. \cite{moynihan_utilization_2014} surveyed 23 steel framed buildings in the UK and found an average utilization rate (design load of member divided by the member capacity) of less than 50\%, indicating significant unnecessary material consumption. Of the 23 buildings, the designers of 8 noted fabrication and construction efficiency as the primary driver for oversizing members. 

When project circumstances permit the realization of complex geometries, shape rationalization is often performed to reduce fabrication and construction challenges while maintaining architectural intent. In the case of free-formed structural envelopes and facades, much of the focus is in converting arbitrary surfaces to be either developable---surfaces with 0 Gaussian curvature that can be flattened without distortion, or ruled---surfaces that can be made of  linear elements.  \cite{pottmann_architectural_2015} provides an overview of many of the challenges and solutions towards this discretization, and notes the existing barriers to make any technique completely generalizable. 

Alternatively, designers can limit themselves with tools that only generate developable and ruled surfaces to ensure their initial intent can be readily realized. The hyperbolic paraboloid concrete shell structures of Felix Candela were based on ruled surfaces whose formwork can be assembled with standard  dimensional lumber \citep{candela1955structural}; \cite{glaeser2007developable} provides a computational method for the same effect for developable surfaces.

Complexity in design is not always considered detrimental. During the option evaluation phase of design, complexity can be interpreted as \textit{diversity}. \cite{brown_quantifying_2019} investigated multiple metrics of diversity in the context of parametric design exploration. The values of each design-changing parameter are represented as components of a point in $\mathbb{R}^n$, with different points in the parameter space representing different designs. Diversity is then measured by the spread of the design points in parameter space.

Rationalization for building efficiency is not restricted to geometry. \cite{stephen2018clustering} considered the domain of regional load demands across the United States as the basis of rationalization. By clustering similar load demand regions, material-efficient emergency shelters can be designed for a reduced set of loading requirements, rather than a single oversized shelter. 

For topology optimized truss structures, methods of integrating construction complexity based on unit node and element costs have been developed by \cite{asadpoure2015incorporating} and \cite{torii2016design}. These methods seek to minimize the accumulation of lower-impact nodes and elements in the ground-based topology optimization approach.

For the rationalization of spatial truss nodes, \cite{koronaki_rationalization_2020} pursued an alternative method of characterizing nodal demands for the purpose of complexity reduction. A best-fit reference plane is placed among normalized element vectors acting at a node using Principal Component Analysis, and a consistent reference element is selected. All other elements are characterized by the relative angle to this reference element with respect to the reference plane. Dissimilarity between two nodes is measured by the difference in the ordered set of element angles. The nodal positions are then optimized to maximize the number of similar nodes within a set angle tolerance.

\subsection{Shape similarity}\label{subsec:2_2}
By representing forces as a three dimensional shape, we leverage shape recognition tools in the field of computational geometry. Shape similarity plays a key role in fields ranging from molecular biology to 3D animation. Applications include indexing from expansive catalogs of parts to feature recognition and shape interpolation for character modeling. \cite{biasotti2016recent} provided a state of the art overview of a wide range of proposed shape similarity descriptors. The output for most shape descriptors is the feature vector, a fixed-length vector representation of a given shape. Dissimilarity between two shapes is measured by the distance between feature vectors.

Two general categories of shape descriptors exist: geometry-based and topology-based. For geometry-based shape similarity, the exact positions of shape vertices, edges, and faces are analyzed and compared. The challenge for these methods is controlling invariance, where the same shape in different orientations is not inherently captured by the analysis.  \cite{chen_visual_2003} used a series of evenly spaced virtual cameras around a given 3D object to find 2D projection images from all angles. An optimal alignment procedure, akin to rigidly rotating the positions of the virtual cameras, is performed to compare two shapes with minimum distortion. Advancements in view-based shape recognition include hypergraph representations of the captured 2D images \citep{wang_3d_2016}, and various learning-based processes, summarized by \cite{qi_review_2021}.

Topology-based methods consider the underlying connectivity of vertices and edges of the input shape as the basis of shape characterization; these methods exploit intrinsic properties of the topology that are independent of reference frames and orientations. The Heat Kernel Signature, the intrinsic characterization of simulated heat diffusion in a meshed object, has been used to recognize shapes in varying non-rigid deformations \citep{bronstein2010scale,raviv2010volumetric}, and for recognizing highly detailed features on a specific shape \citep{bronstein_shape_2011}.

A unique group of shape descriptors are those that exploit orthonormal basis functions on the circle (2D shapes) and sphere (3D shapes). These methods are not topology-based, but have intrinsic invariance to object orientation. In general, the input shapes are transformed into periodic circular or spherical functions, that are then expanded as sums of basis functions using Fourier series (2D) or spherical harmonics (3D). the unique coefficients of the expanded functions allow for a fixed-length vector representation of the shape they represent. 2D shape recognition using Fourier expansion was first proposed by \cite{persoon1977shape}. The contour of an input shape is represented as a periodic function, which is then represented as a weighed sum of cosine functions of increasing frequency; the scalar coefficients of each term provides the rotation-invariant feature vector. The use of Fourier shape descriptors have been expanded for finer resolution of more complex shapes \citep{zhang_shape-based_2002,yang_multiscale_2019}, as well as for shapes with holes for spatial geography applications \citep{xu_shape_2017} .

The 3D equivalent is the decomposition of the input shape into its expanded spherical harmonics. The spherical harmonic shape descriptor, first proposed by \cite{kazhdan2003rotation}, intersects a given shape with a series of concentric spherical shells. The points of intersection for a given shell represent the shell's spherical function, which can then be decomposed as a weighed sum of spherical harmonic basis functions; by taking the $L_2$-norm of the frequency components of each expansion series, a rotation-invariant characterization of the given shape can be determined. This inherent rotation invariance is the key to consistent characterization and comparisons between shapes, and is the method chosen for adaption in this paper; the mechanics of this process is described in detail in Section \ref{sec:3}. Advancements on this method have been proposed by \cite{wang_3d_2016}, who optimized the expansion coefficients for more complex shapes with detailed feature, and by \cite{esteves2018learning} for applications in convolutional neural networks. 

\subsection{Geometric representation of force}\label{subsec:2_3}
Geometric representation of force for analysis and design is most recognized in the field of Graphic Statics. Developed in its current form in the 19\textsuperscript{th} century, graphic statics represents the axial forces of linear elements of a \textit{form} diagram with a reciprocal \textit{force} diagram \citep{maxwell1864xlv, rankine1864xvii}. In 2D, the reciprocal force diagram represents a loaded linear element with an equivalent line whose length is proportional to the internal force; in 3D, it is represented by a 2D polygon whose area is force-proportional. Nodal equilibrium is represented as a closed polygon (2D) or polyhedron (3D). Although initially used as a procedural method of analysis and design, recent algebraic formalizations of graphic statics principals have allowed computational graphic statics to resurge as a modern tool in design and analysis \citep{van2014algebraic, akbarzadeh20163d, hablicsek2019algebraic, NEJUR2021103003, Mozaffari2021}.

\subsection{Research gap and contributions}\label{subsec:2_4}
In \cite{brown_quantifying_2019}, a quantifiable measure of design variation was presented. However, the method quantifies inter-design variation, rather than variation within a single design, and does not consider internal force magnitudes. We develop a similar metric of design complexity, but focus on the internal force demands of a single design. 

In \cite{koronaki_rationalization_2020}, a method of spatial truss node rationalization was developed. One drawback is the omission of force magnitude when characterizing a structural node. We present an alternative method that includes force magnitudes, and further, does not require a reference frame alignment process for each node.

We adapt the spherical harmonic shape descriptor method by \cite{kazhdan2003rotation} by converting nodal force demands into geometric spherical objects, and extending the resulting shape descriptors to useful applications in the building design, both as a performance metric in the early design phase, and as the basis of clustering for design rationalization. We bypass the concentric shells method and directly convert nodal forces to a single spherical function. Along with reduced computation, it also omits the need to calibrate the number and spacing of these shells to fully characterize a given shape.

Although the equilibrium shapes of graphic statics are the most natural geometric representations of force demand, current computational tools for the analysis of spatial truss structures are limited to self-stressed networks or specific typologies. Further, using polygons and polyhedra requires either secondary processing to ensure rotation invariance during comparison, or the concentric shell method if using spherical harmonic descriptors. We present an alternative geometric representation of force that streamlines inter-nodal comparisons.

\begin{figure}[h]
    \centering
    \includegraphics[width = \textwidth]{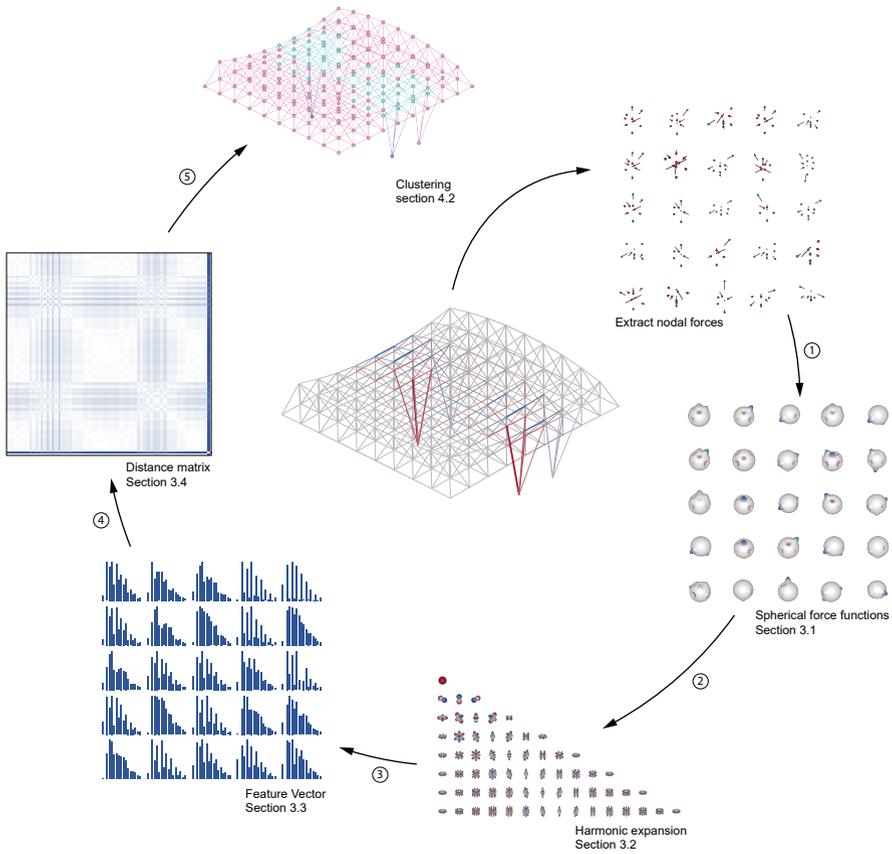}
    \caption{Methodology overview: 1. the nodal force demands are converted to spherical functions that encode force magnitudes and orientations, 2. the force functions are expanded into spherical harmonic basis functions, 3. the feature vector is extracted from the harmonic decomposition, 4. the distance matrix between all feature vectors is determined, 5. clustering analysis and complexity metrics are enabled by the distance matrix.}
    \label{fig:intro}
\end{figure}

Our methodology is summarized in Figure \ref{fig:intro}.

\section{Methodology}\label{sec:3}
\begin{figure}[h]
    \centering
    \includegraphics[width = 0.5\textwidth]{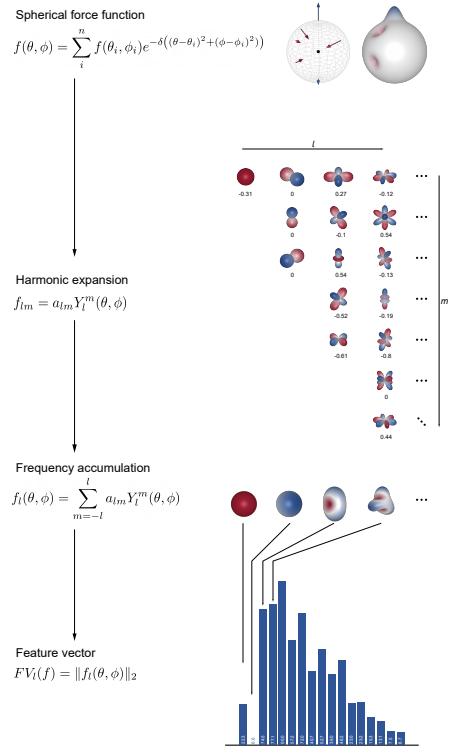}
    \caption{Overview of the conversion of nodal force demands to a rotation-invariant fixed-length feature vector.}
    \label{fig:overview}
\end{figure}

This section provides a detailed overview of the conversion of an arbitrary spatial truss node and its force demands to a rotation-invariant feature vector using the spherical harmonic shape descriptor developed by \cite{kazhdan2003rotation}. We modify the process to be readily applicable for spatial truss analysis by providing a deterministic method of converting nodal force demands to a single spherical force function, and calibrate the resolution of the feature vector to ensure the force function is well represented. Multiple visualization methods are also developed to provide designers with rapid insight on nodal complexity for a given spatial truss form. The process is summarized in Figure \ref{fig:overview}.

The examples provided in this section are based on the structure in Figure \ref{fig:problem}; multiple variations of this spatial truss are also analyzed for comparison throughout this paper. All variations contain the same topology, number of nodes and elements, and overall enclosed planar area. Linear elastic analysis is performed to extract all internal forces. In total, 100 variations were generated using Latin hypercube sampling of a parameterized model \citep{danhaive2021design}.  The model has six design variables which correspond to the z-coordinates of control points of a bilaterally symmetric NURBS surface used to define the upper curved geometry of the two-layer trussed roof.  In this model, the bottom layer and positions of the columns and supports remain constant.  The parametric model is intended to generate a range of designs of varying efficiency achieved through structural morphology, as a proxy for design alternatives that would be considered in early-stage structural exploration.

\subsection{Spherical function representation}\label{subsec:3_1}
The spherical harmonic shape descriptor was initial developed to be generalizable to any 3D object, where the intersection of spheres of increasing radius with the given object creates binary spherical functions. This step takes added computation to sample the intersection points, and requires calibration of the number of spheres required to capture small details. We bypass this step by directly creating a single representative spherical function that captures all information of nodal force demand. 

We first convert the set of forces acting at a node to its equivalent function by considering the spherical coordinates of entry (for compression forces) and exit (for tension forces). At each coordinate, the scalar function value is equal to the respective axial force magnitude. To provide the smooth, square-integrable function required for spherical harmonic expansion, we convert the singular axial force values as Gaussian distributions on the surface of the sphere:

\begin{equation}
    f_i(\theta,\phi) = f(\theta_i, \phi_i)e^{-\delta\left((\theta-\theta_i)^2 + (\phi - \phi_i)^2\right)}
\end{equation}

Where $f(\theta_i, \phi_i)$ is the magnitude of force $i$ acting at $\theta_i, \phi_i$. We take the physics convention of $\theta$ as the angle from the polar axis $z$, and $\phi$ as the angle from the $x$ axis in the $xy$ plane. Because the end result of the spherical harmonic shape descriptor is invariant to rotation, a consistent reference axis is not required when determining the spherical force function, but is taken as the global XYZ axes for convenience. For a node with $n$ forces, the spherical force function is then the sum of all Gaussian force distributions:

\begin{equation}
    f(\theta,\phi) = \sum_i^n f_i(\theta, \phi)
\end{equation}

\begin{figure}[h]
    \centering
    \includegraphics[width = 0.5\textwidth]{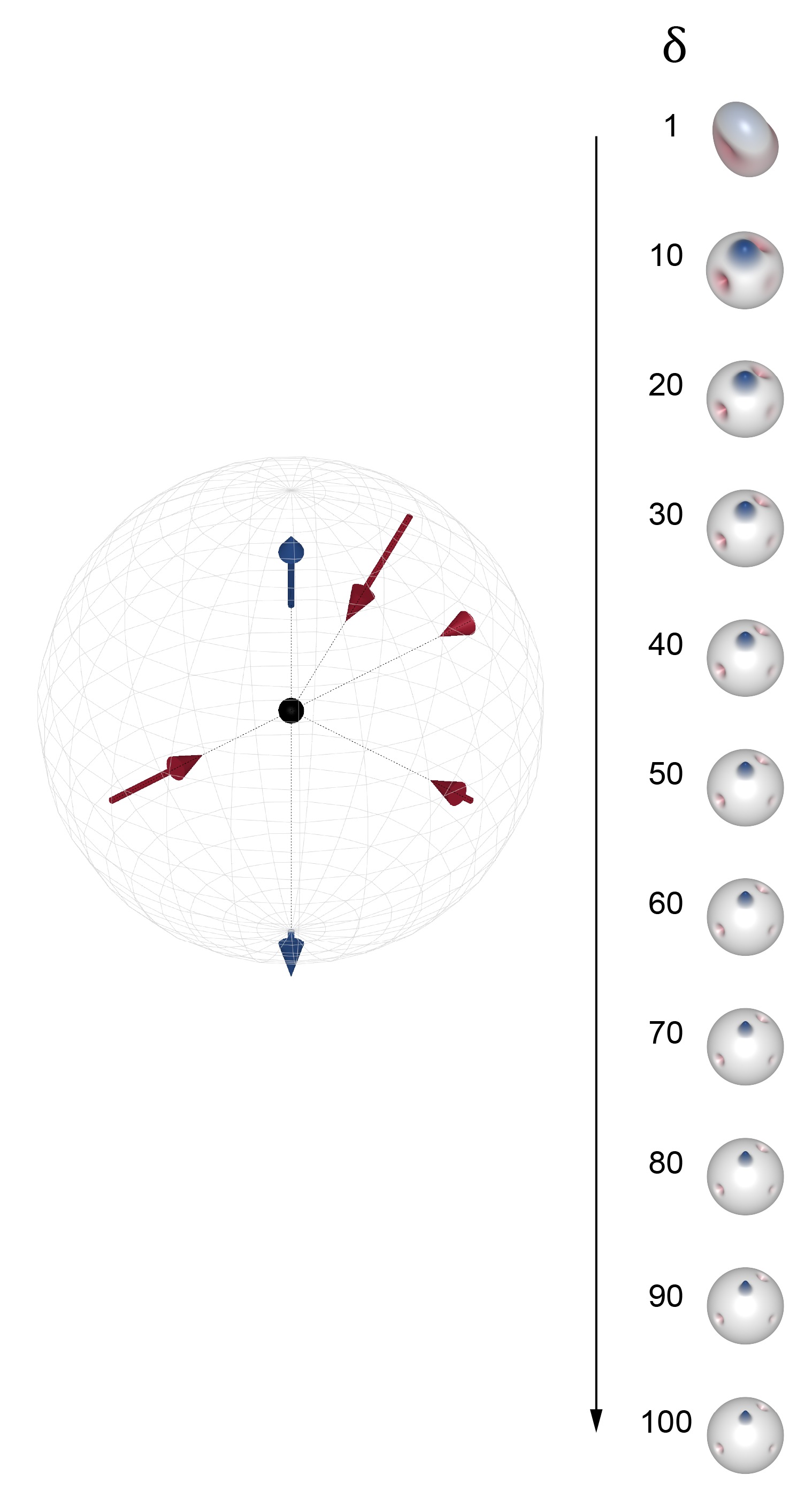}
    \caption{Spherical functions of varying values of $\delta$. We take $\delta=20$ for subsequent analyses for good representation of all force sets while remaining computationally efficient.}
    \label{fig:delta}
\end{figure}

The selection of the variation factor $\delta$ requires calibration. Large values approaches a dirac delta function of a singular spike at the location of the force, resulting in poor approximations when expanding into spherical harmonics; small values smooth the forces over large regions of the sphere, and result in the loss of individual force delineation. We show a range of values for $\delta$ in Figure \ref{fig:delta}. We take $\delta = 20$ for all subsequent analyses for good delineation between close force vectors and good approximation by spherical harmonic expansion (Section \ref{subsec:3_2}). Structures with closely packed elements at nodes may require a larger value for better individual force delineation.

\begin{figure}[h]
    \centering
    \includegraphics[width = 0.5\textwidth]{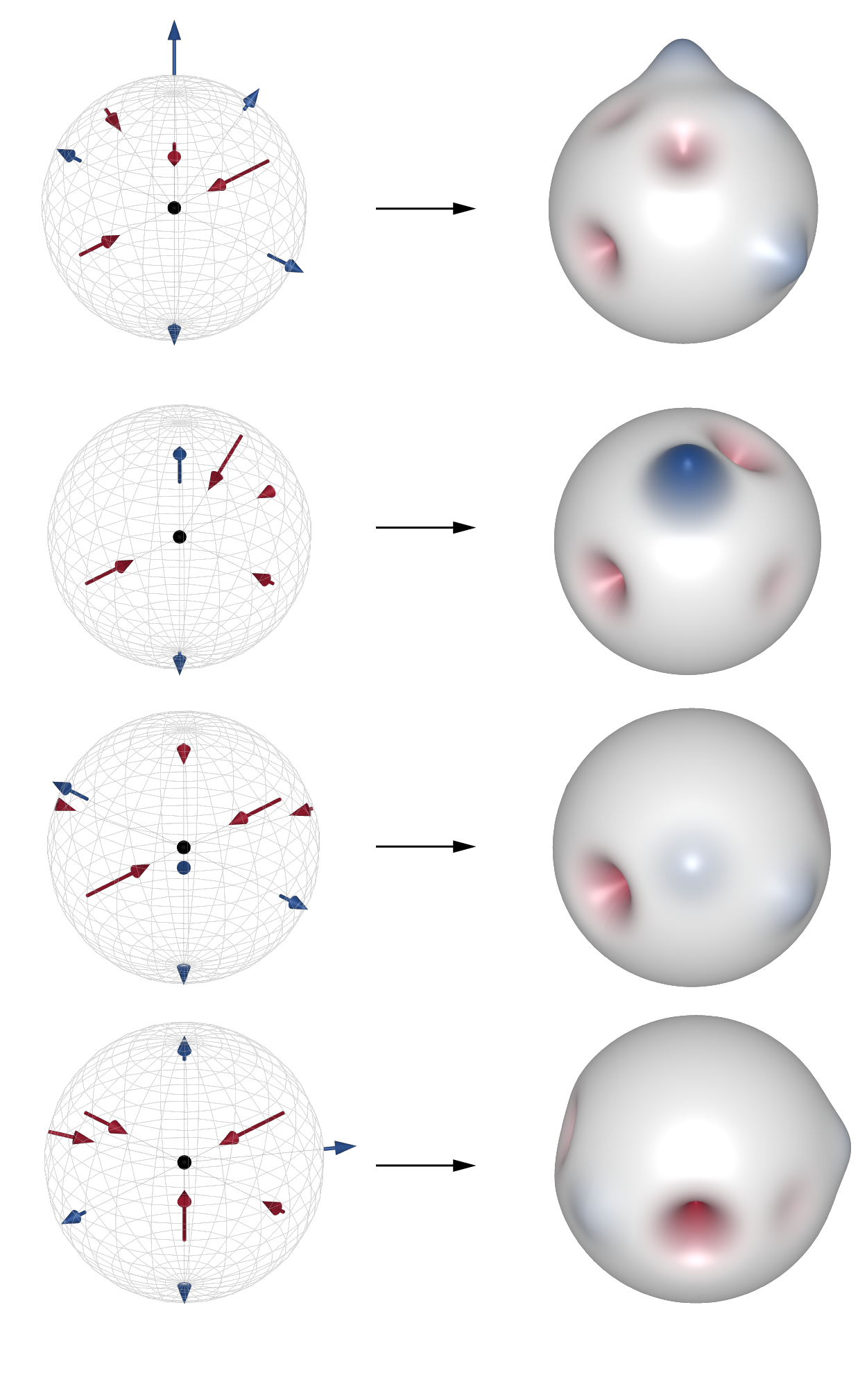}
    \caption{Spherical force functions for varying nodal force demands.}
    \label{fig:forces}
\end{figure}

The conversion of force demands into spherical functions is performed for all nodes of the structure; a sample is provided in Figure \ref{fig:forces}. The depth of the indentations and protrusions are proportional to the forces within each node, but are not to scale when comparing two nodes. This deformation of the spherical surface is intended for better visualization, but the force function itself remain scalar-valued one-to-one functions on the surface of the unit sphere, $S^2$.

\subsection{Spherical harmonic expansion}\label{subsec:3_2}
\begin{figure}[h]
    \centering
    \includegraphics[width = 0.5\textwidth]{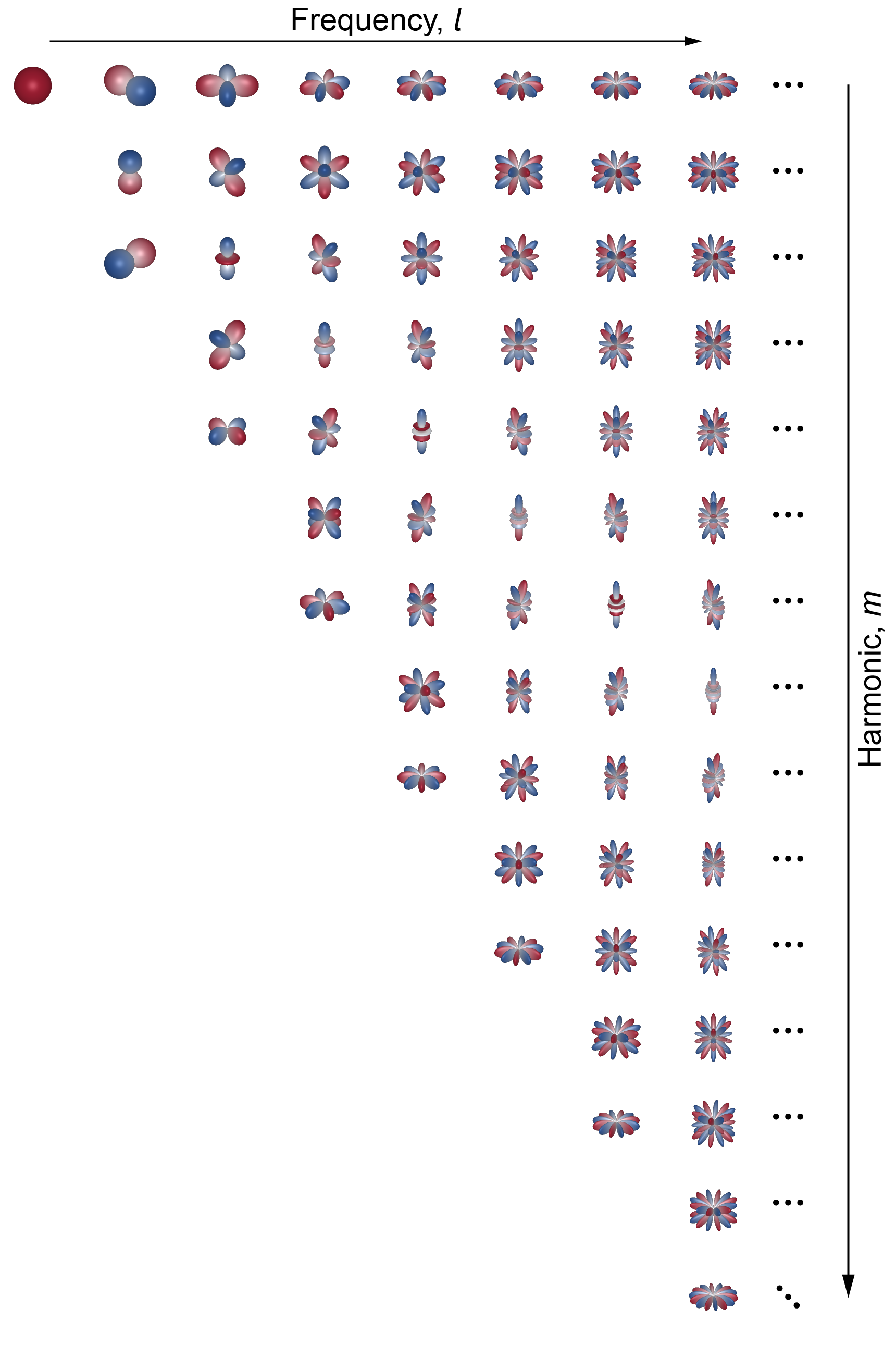}
    \caption{Spherical harmonics for $l \in [0,7]$; $m \in [-l, l]$}
    \label{fig:harmonics}
\end{figure}

With suitable spherical representation of nodal force demands, we then decompose each function into a weighed sum of the spherical harmonic basis functions. The spherical harmonics are the eigenfunction solutions to Laplace's equation on the sphere: $\nabla^2f(\theta,\phi) = 0$, where $\nabla^2$ is the Laplace Operator, the divergence of the gradient of the scalar function $f$. The solutions, denoted $Y_l^m(\theta,\phi)$, are grouped into frequencies (degree), $l$, and harmonics (mode), $m$, and form a complete orthonormal set of basis functions on the sphere. Frequencies span from zero to infinity in integer increments, with each frequency containing $2l+1$ modes that span from $[-l,l]$; the spherical harmonics of the first 8 frequencies and their corresponding harmonics are shown in Figure \ref{fig:harmonics}, with red regions indicating negative function values and blue regions indicating positive values. The shapes are determined by scaling the radial component of the spherical function, $r$, by the magnitude of the function value.

As the spherical harmonics form a complete orthonormal set of basis functions, any square-integrable function $f(\theta,\phi)$ can be represented as a summation of these harmonics by:

\begin{equation}
    f(\theta,\phi) = \sum_{l=0}^{l_{max}} \sum_{m=-l}^l a_{lm}Y_l^m(\theta,\phi)
    \label{eq:expansion}
\end{equation}

Where $a_{lm}$ is the associated scalar coefficient specific to the input function $f$. As $l_{max} \rightarrow \infty$, an identical representation of the input function is determined. The precision of the harmonic expansion of a spherical function depends on the resolution of $l_{max}$ taken by the summation in Equation \ref{eq:expansion}. The relative error of this resolution for a given spherical force function $f$ is taken as:

\begin{equation}
    \text{error}_{l_{max}} = \frac{\lVert f - \sum_{l=0}^{l_{max}} \sum_{m=-l}^l a_{lm}Y_l^m \rVert_2}{\lVert f \rVert_2}
    \label{eq:error}
\end{equation}

\begin{figure}[h]
    \centering
    \includegraphics[width = \textwidth]{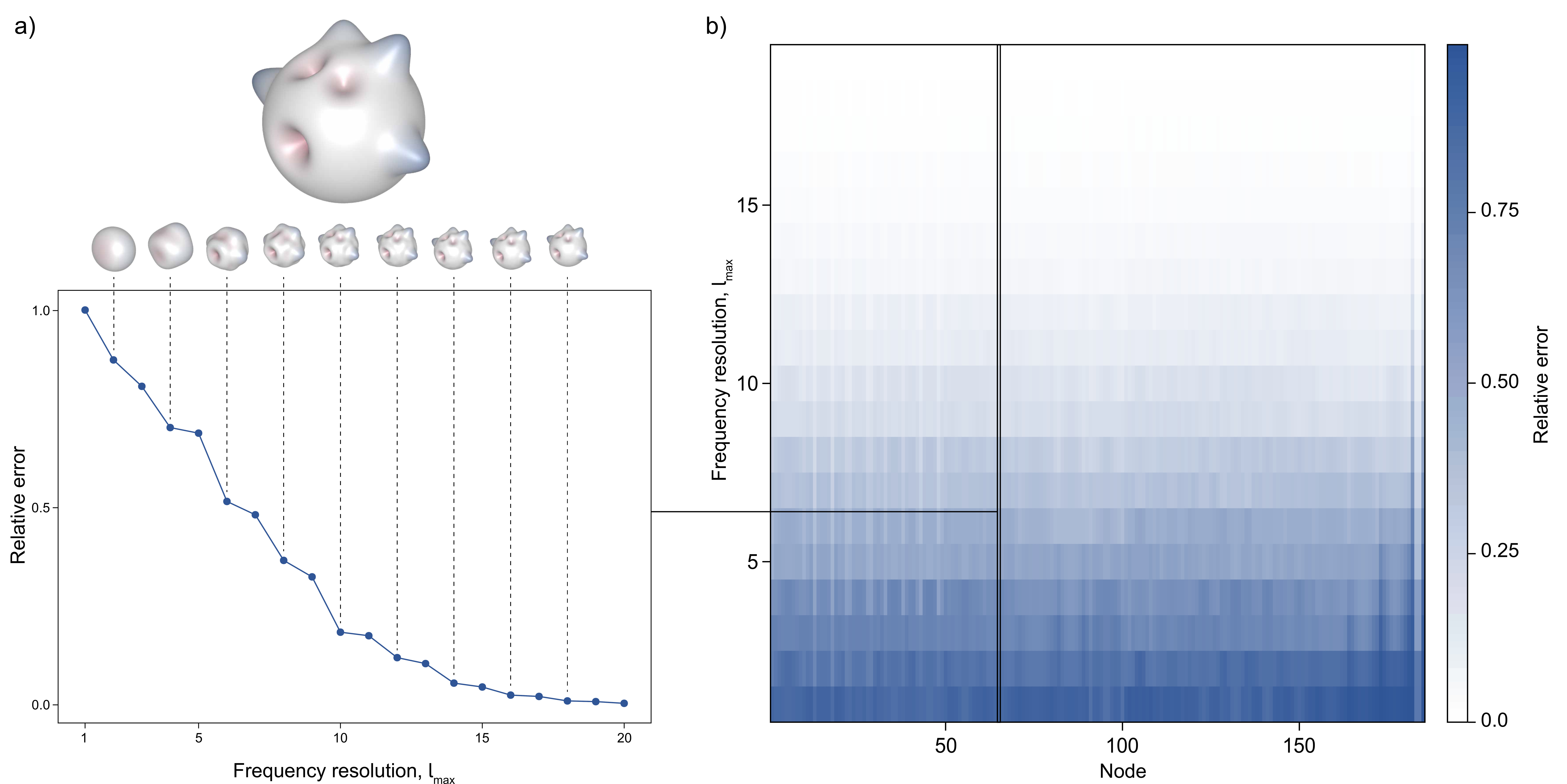}
    \caption{The effect of $l_{max}$ on the effective representation of a spherical force function using spherical harmonic expansion: a) for a single node, b) for 185 nodes.}
    \label{fig:errors}
\end{figure}

We expand a representative nodal force function to varying degrees of $l_{max}$ and observe the resolution error in Figure \ref{fig:errors}a. Steady reduction in error is observed with increasing resolution, especially in the early stages, with a slower convergence to zero past $l_{max} \approx 14$. We perform this error analysis on all 185 nodes of the example structure in Figure \ref{fig:errors}b, and observe similar frequency-error relationships. We take $l_{max}=16$, or the first 17 frequencies, for all subsequent analyses, with a mean relative error of 1\%.

There is a circular relationship between the variance factor $\delta$ used in the force function representation of nodal demand, and the error of approximation when choosing $l_{max}$. Larger values of $\delta$ will sharpen the positions of the force values and require more terms during the spherical harmonic expansion to approximate its shape. As noted in Section \ref{subsec:3_1}, the value of $\delta$ will require verification depending on the design of the specific spatial truss structure, and accordingly, $l_{max}$ may have to be adjusted. Our selection of $l_{max} = 16$ equals the value taken in the original method by \cite{kazhdan2003rotation}. 

\subsection{Frequency accumulation and feature vector representation}\label{subsec:3_3}
We continue to follow the spherical harmonic shape descriptor process by summing the weighed basis functions in each frequency, $l$, to create a single spherical function for each frequency:

\begin{equation}
    f_l = \sum_{m=-l}^l a_{lm}Y_l^m
\end{equation}

This summation provides the key characteristic of rotation-invariance. The set of all harmonic functions within a fixed frequency, $Y_l^{-l}, ..., Y_l^l$, is a representation of the rotation group $SO(3)$. Any partial feature on the input function that is represented by the specific set of harmonic functions of frequency $l$ is represented by the same set of functions irrespective of its orientation in space. Although the corresponding coefficients of the harmonic functions $a_{lm}$ may be rearranged depending on the orientation of the reference axes, the overall summation within the frequency does not change. As such, the overall contribution of a given frequency in representing a shape can be defined as the $L_2$-norm of the summed function, or its \textit{energy}. This scalar positive energy value then creates the components of the rotation-invariant feature vector: 

\begin{equation}
    FV(f) = \left[ \lVert f_0 \rVert_2, ..., \lVert f_n \rVert_2 \right]
\end{equation}

The feature vector is the final outcome of the conversion process, and the arbitrary sets of forces acting on different structural nodes can now be compared. 

\subsection{Measuring dissimilarity}\label{subsec:3_4}
By expanding all nodal functions to the same value of harmonic frequencies, a fixed-length feature vector representation of each nodal force function is determined. The dissimilarity between two nodes is captured by the $L_2$-norm of the difference between their respective feature vectors:

\begin{equation}
    d(\text{node}_1,\text{node}_2) = \lVert FV_1 - FV_2 \rVert_2
\end{equation}

\begin{figure}[h]
    \centering
    \includegraphics[width = 0.5\textwidth]{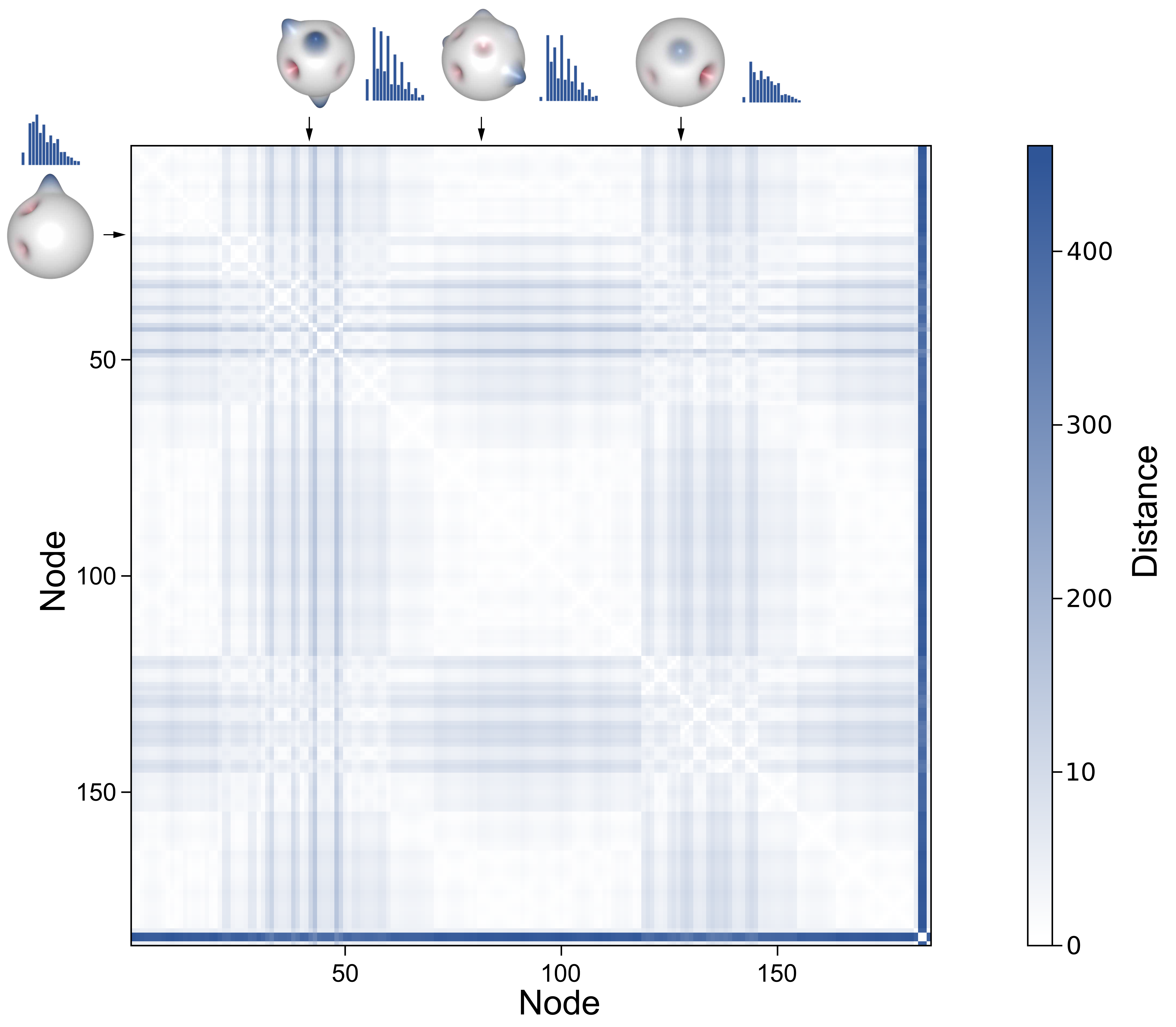}
    \caption{Distance matrix of all nodes. Darker regions indicate nodes that are dissimilar from each other.}
    \label{fig:costmap}
\end{figure}

This distance can be measured between any two nodes, regardless of the number of connected elements and its position on the structure, and represents both force orientations and magnitudes. By taking all inter-nodal distances within a structure, a symmetric distance (or dissimilarity) matrix is created, as seen in Figure \ref{fig:costmap}.

The distance matrix provides a visual overview of the variation of nodal force demands across the structure. Rows (or columns) with consistently large distances indicate connections with highly unique force demands, and are ideal targets for more intensive connection design and detailing. The dark bands of consistent dissimilarity in Figure \ref{fig:costmap} represent the four support connections at the base of the example structure. An alternative visualization to represent the dissimilarity of structural nodes is through classical multidimensional scaling (MDS) of the $n$-dimensional feature vectors into parsable lower dimensions. Classical MDS is a dimension reduction technique that embeds higher dimensional points into an arbitrarily lower dimension while best preserving the initial distance matrix \citep{torgerson1952multidimensional}. 

\begin{figure}[h]
    \centering
    \includegraphics[width = \textwidth]{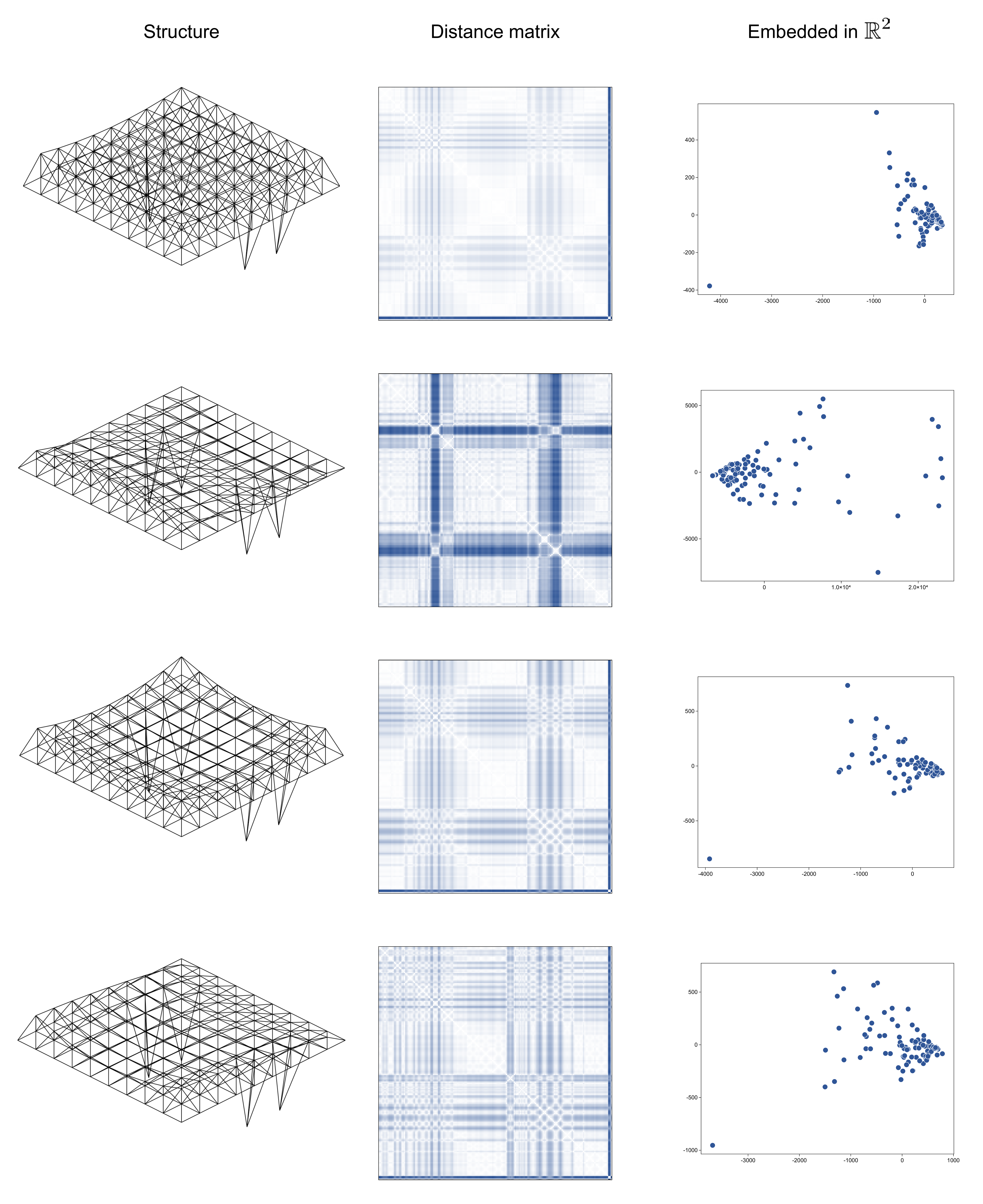}
    \caption{Distance matrix and 2D embedment of feature vectors using MDS for four design alternatives.}
    \label{fig:mds}
\end{figure}

An example of how the distance matrix can be used for visual understanding of nodal force demand variation, as well as a Classical MDS embedment in $\mathbb{R}^2$, is shown in Figure \ref{fig:mds}. Four designs with the same topology, planar dimensions, and materials as in Figure \ref{fig:problem} are shown along with their distance matrices and the 2 dimensional embedment of their nodal feature vectors. The absolute coordinates of the embedded points hold no significance, but the distances between them are representative of the Euclidean distances in $\mathbb{R}^n$. By visual inspection, it is evident that designs with greater inter-nodal dissimilarity have more dark regions in the distance matrices, as well as a greater dispersion of embedded feature vector points. During the project ideation phase, designers can look to both the distance matrix and embedded point visualizations of the nodal feature vectors to indicate designs that are better or worse suited for either mass customization or standardization of the joints. Larger dispersions of the embedded points (or darker distance matrices) indicate a greater penalty in standardization.

\begin{figure}[h]
    \centering
    \includegraphics[width = 0.5\textwidth]{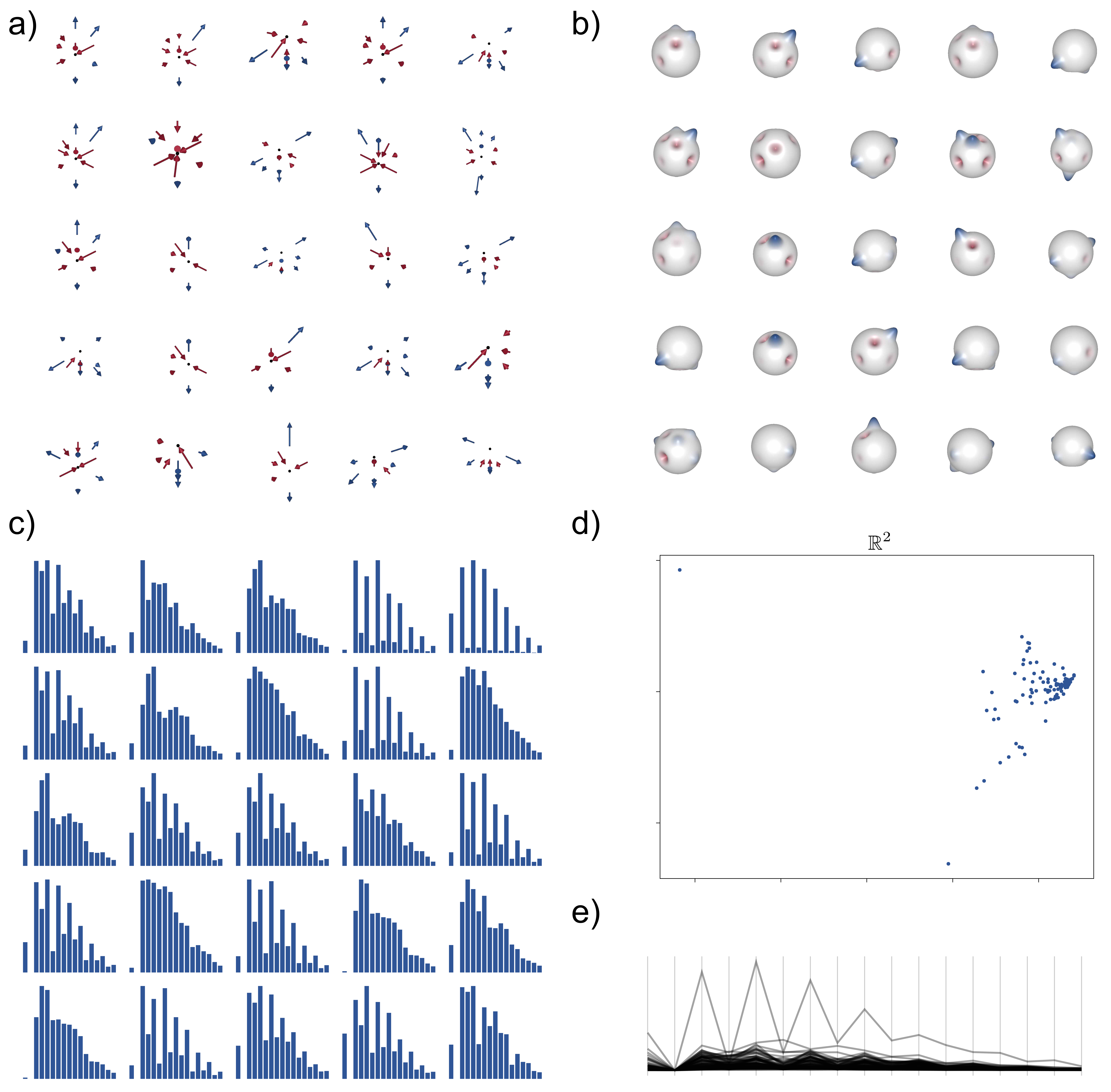}
    \caption{Equivalent representations: a) force vectors acting on node, b) spherical force functions, c) feature vectors, d) embedded points in $\mathbb{R}^2$, e) superimposed parallel coordinate plots of feature vectors.}
    \label{fig:equivalents}
\end{figure}

Geometric analysis of forces allow for a wealth of visualizations during each stage of the analysis process. We show 5 equivalent representations of nodal force demands in Figure \ref{fig:equivalents}, each useful for varying stages of analyses or visual understanding of force demand variance. We introduce the superimposed parallel coordinate plot of feature vectors in \ref{fig:equivalents}e as an alternative to the MDS embedded points as a method of visualizing the degree of similarity between nodes. The visually distinct line group in the parallel coordinate plot again represent the unique force demands acting at the compression supports in Figure \ref{fig:problem}.

\subsection{Characteristics of the feature vector}\label{subsec:3_5}
The feature vector representation of nodal force demands has two key characteristics. First, the feature vectors scale linearly to the input function:

\begin{equation*}
    FV(cf) = cFV(f)
\end{equation*}

\begin{figure}[h]
    \centering
    \includegraphics[width= 0.5\textwidth]{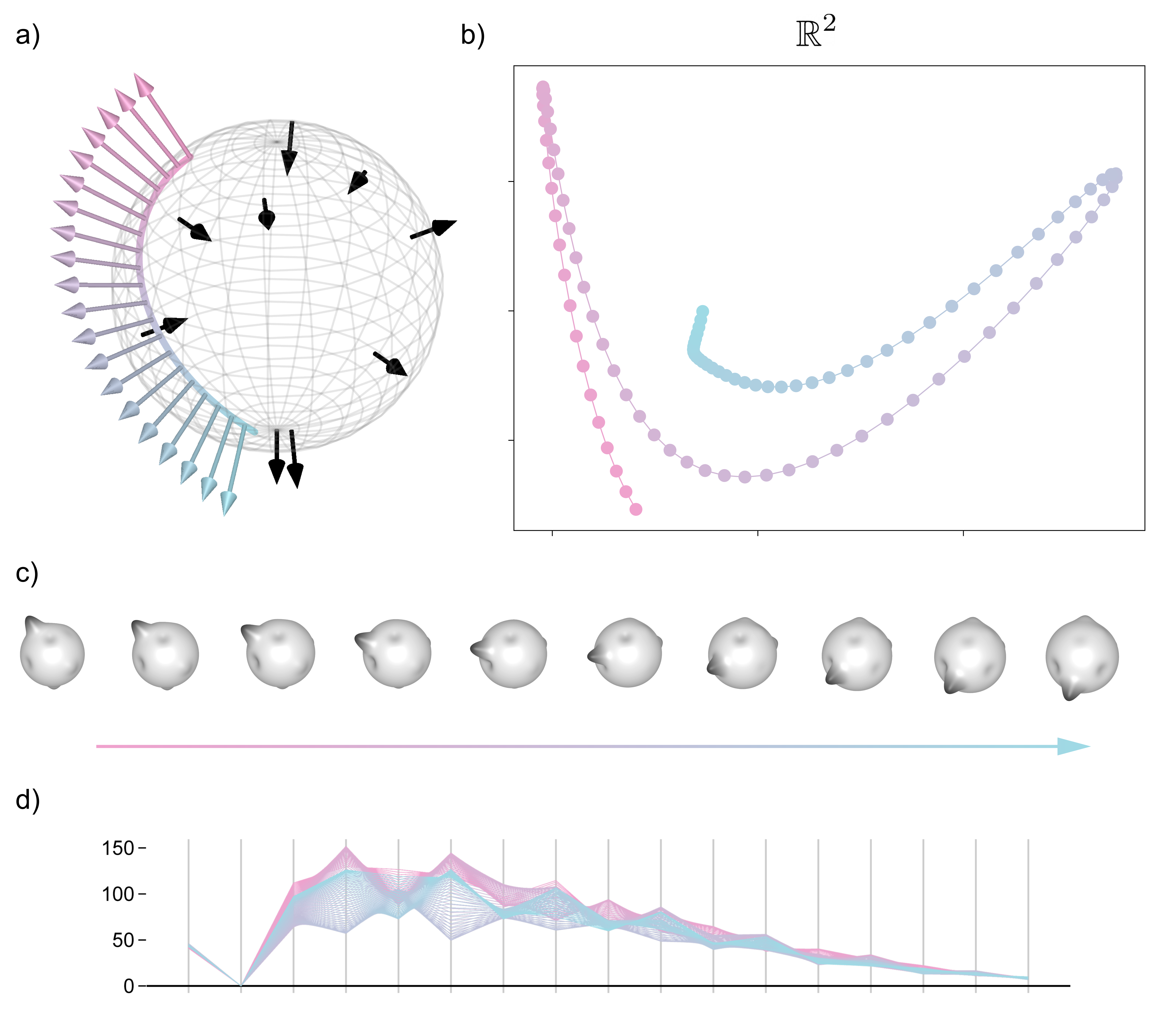}
    \caption{Smooth changes of input: a) the transition path of a single force in a nodal force set, b) the resulting positions of the feature vectors embedded in $\mathbb{R}^2$ using MDS, c) the spherical force functions of the altered force directions, d) parallel plot of changing feature vectors}
    \label{fig:movingforces}
\end{figure}

When only member orientations are of concern when calculating inter-nodal dissimilarity, force demands can first be normalized before harmonic expansion and analysis. Second, we observe that smooth variations of the input spherical force functions result in smooth changes in the output feature vector. From a representative node and its associated forces, we move one force along an arc (shown in Figure \ref{fig:movingforces}), and plot both its feature vector in parallel coordinate form and its relative $\mathbb{R}^2$ embedded position to all other sampled force positions. All other forces acting on the node remain at the same position, but their magnitudes are changed at each iteration to maintain equilibrium. We observe small changes in the feature vector with small changes of the force position, best observed by the incremental distances between the embedded feature vector points in Figure \ref{fig:movingforces}b. This gradual increase is smooth, but not predictable; the start and end positions of the moving force vector have greater similarity than regions in between. 

Third, as normed frequency \textit{energies} are taken as the components of the feature vector, they are invariant to direction of the force acting at the node:

\begin{equation*}
    FV(f) = FV(-f)
\end{equation*}

This represents a loss of information once a feature vector is calculated for a given node, resulting in a non-bijective map from nodal force demands to the space of feature vectors. It is not possible to invert the feature vector to the input distribution of compression and tension forces. This limitation means that tasks such as interpolating new force vector sets is not possible directly, but could be achieved through iterative optimization methods in future work.

Finally, it is emphasized that the components of the feature vector do not directly correspond to a specific force acting on a node. Rather, they represent overall regional features of the force function captured by a given frequency. A large component in the feature vector should not be interpreted as a particularly large, singular force. However, this lack of direct correspondence between the feature vector and its associated node is key to its function: any truss node, with an arbitrary number of elements and orientations, can be represented by a vector of the same dimension.

\subsection{Methodology summary}\label{subsec:3_6}
The procedure to characterize and compare a set of structural nodes and their force demands require the following steps: 1. conversion into a spherical force function, $f(\theta, \phi)$, created as a sum of Gaussian distributions of individual force components on the surface of the sphere; 2. expansion into spherical harmonic basis functions; 3. the summation and $L_2$-norm of the frequency energies, $FV_l(f) = \lVert \sum_{m=-l}^l a_{lm}Y_l^m \rVert_2$. This feature vector is of dimension $n=l_{max}$, chosen by the user, and taken by the authors as $16$. Each feature vector can be considered as points in $n$ dimensional space, and the Euclidean distance between any two feature vectors are representative of the dissimilarity between the force demands of their respective structural nodes, both in magnitude and orientation. 

\section{Applications in spatial truss design}\label{sec:4}
We present two immediate applications of the feature vector representation of nodal force demands and its corresponding distance matrix. First, as a performance metric of design \textit{complexity}, measured in proxy as the magnitude of distribution of the feature vectors. Second, as the basis of clustering analysis for strategic reduction of unique connections that are designed and manufactured.

\subsection{Complexity and structural performance}\label{subsec:4_1}
When evaluating multiple design options, a metric of complexity may help reduce design, fabrication, and construction costs. In the case of spatial truss nodes, designers may seek to limit the variations in nodal force demands if limited to a single standardized connection. Conversely, designers should be informed when a large standardization penalty would incur, and suggest the selection of alternative designs. 

By considering the nodal feature vectors as embedded points in $\mathbb{R}^{17}$, we take the radius of the minimal bounding hypersphere as the metric of nodal dissimilarity within a given spatial truss. Comparing the size of the bounding hyperspheres from different designs provides a numeric measure of node complexity, and provides an additional performance metric to work in tandem with structural performance. We note this measure of complexity should only be used comparatively, as the feature vectors do not have direct physical meaning. 

\begin{figure}[h]
    \centering
    \includegraphics[width = \textwidth]{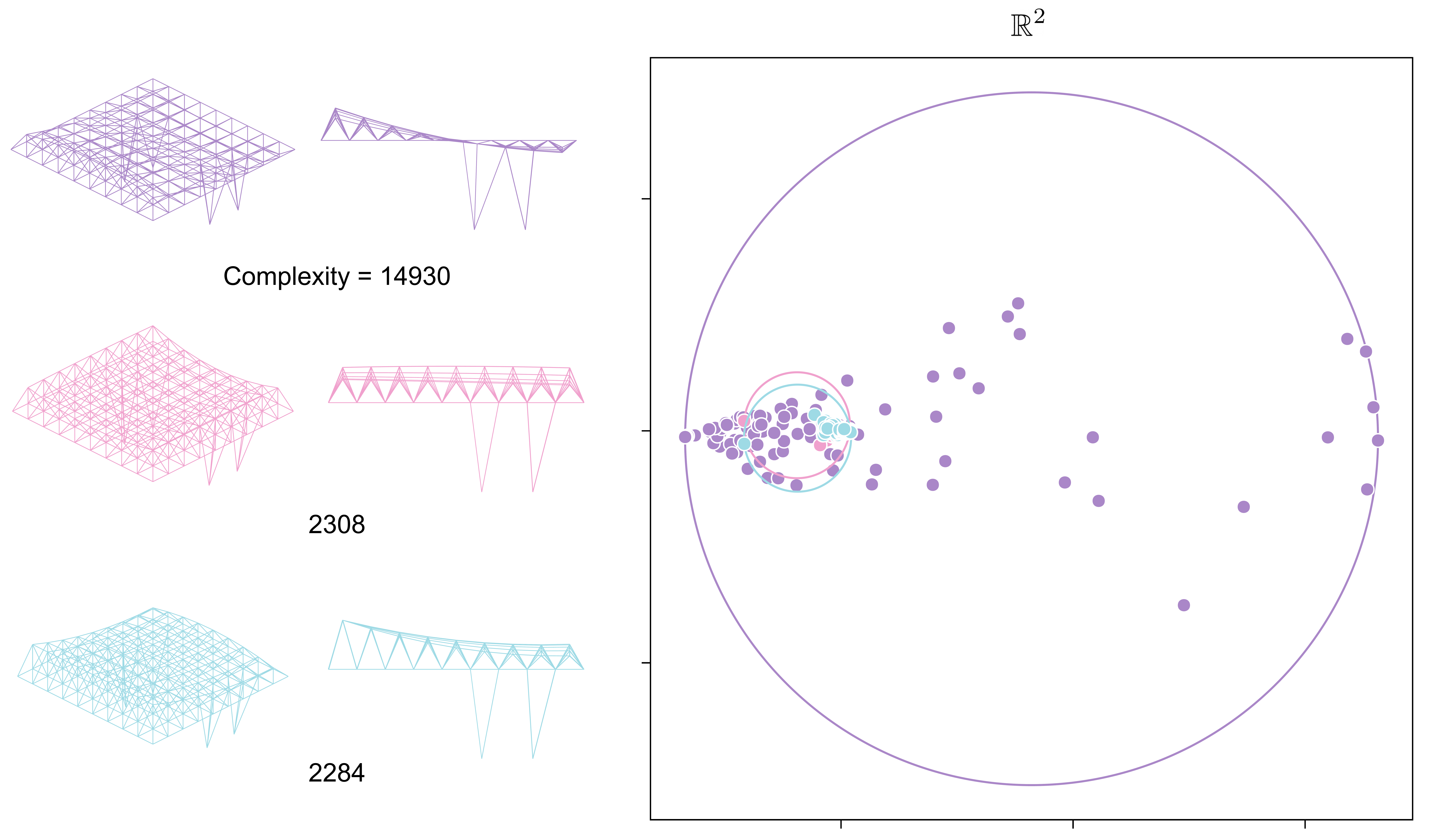}
    \caption{Comparing three alternative designs with their nodal complexity scores, measured by the radius of the minimal bounding hyperspheres of the nodal feature vectors in $\mathbb{R}^{17}$}
    \label{fig:spheres}
\end{figure}

A comparison of three alternative designs, as well as their 2D embedded points and minimal bounding circle, is shown in Figure \ref{fig:spheres}. In the follow examples, the radius of the 2D minimal bounding sphere is exactly the radius of the bounding sphere in the full feature vector space ($\mathbb{R}^{17}$). We observe significant variation in the complexity scores of each design, despite equal numbers of nodes and elements and overall enclosed area. Designers seeking to minimize standardization penalties for connections or who seek to create a reduced number of unique joints should seek designs with smaller bounding circles.

\begin{figure}[h]
    \centering
    \includegraphics[width = \textwidth]{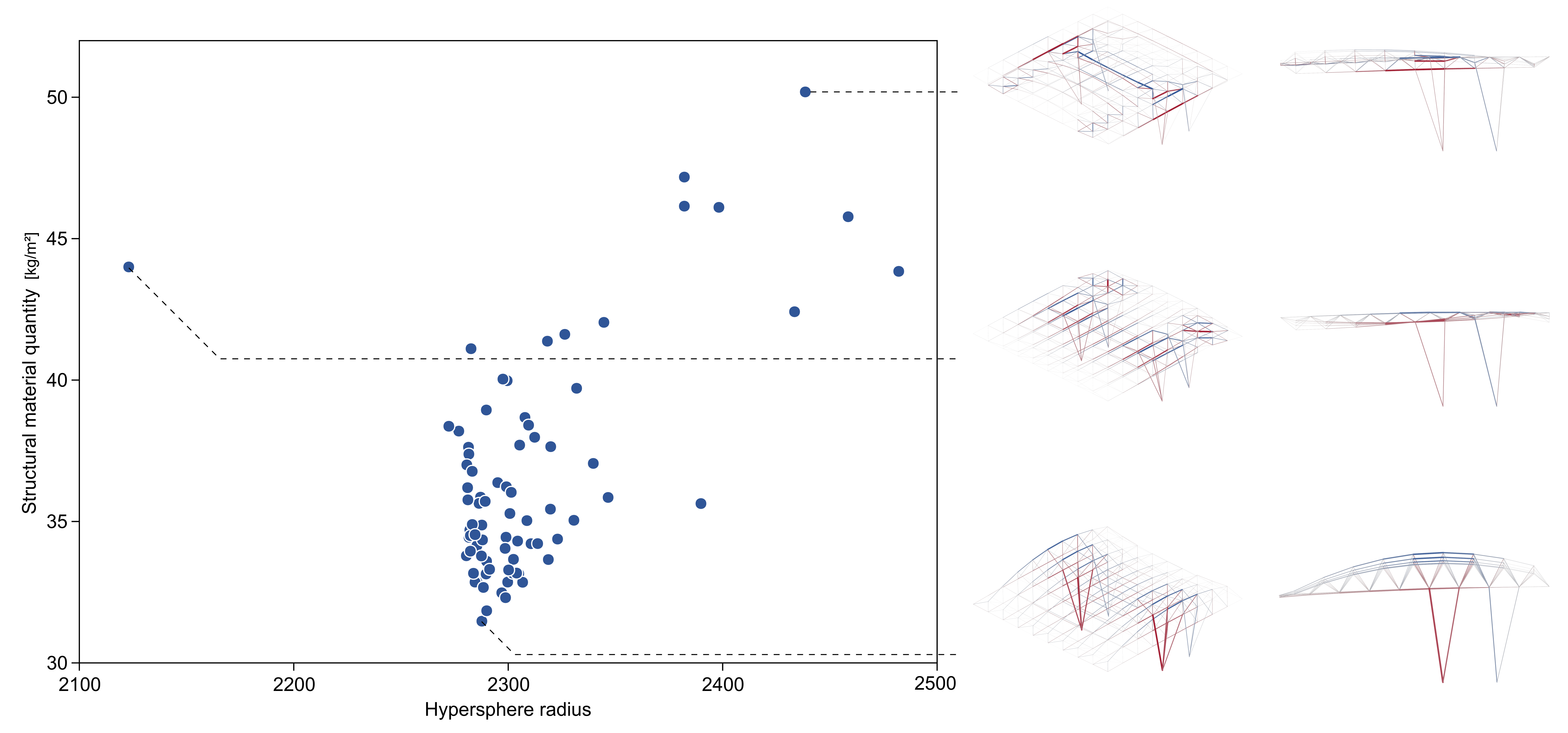}
    \caption{Bi-objective plot of structural material quantity vs. hypersphere radius for 100 spatial truss designs; deflected shapes and internal forces of three key designs highlighted.}
    \label{fig:pareto}
\end{figure}

Node complexity is not the only metric of design performance. More commonly, the global structural performance, measured by the structural material quantity required to withstand the expected loads, drives the design process. We analyzed 100 variations of the initial example structure to investigate the relationship between nodal complexity and structural performance, with the bi-objective plot shown in Figure \ref{fig:pareto}. 

Each structure is given the same initial loading condition and loads as in Figure \ref{fig:problem}. An iterative member sizing process is performed for all structural members by sizing each hollow tube section to withstand its internal load. As cross section properties change the stiffness and force distribution in the structure, this process was repeated until convergence. A minimum cross section area of $400mm^2$ was chosen to represent the smallest standard structural hollow tube section (HSS42.2X3.6). The total mass of this structure was then divided by the enclosure area of 100$m^2$ to provide a normalized score of structural performance. The minimal bounding hypersphere found from the nodal feature vectors of each plot make up the other axis. We highlight the highest performing structure for each objective, as well as a simultaneously poorly performing structure. The bounds of the figure have been adjusted to exclude severe outliers that perform poorly for both objectives, with the largest offender having a hypersphere radii of approximately 15000 and a minimum mass of 150$kg/m^2$. We observe that the design with the best complexity performance does not perform well structurally, but achieves its high degree of nodal similarity due to relatively parallel top and bottom chord planes. The most structurally performant design is more conventional, with a deeper cross section at the support region that tapers towards the cantilever. 

\subsection{Clustering}\label{subsec:4_2}
\begin{figure}[h]
    \centering
    \includegraphics[width = \textwidth]{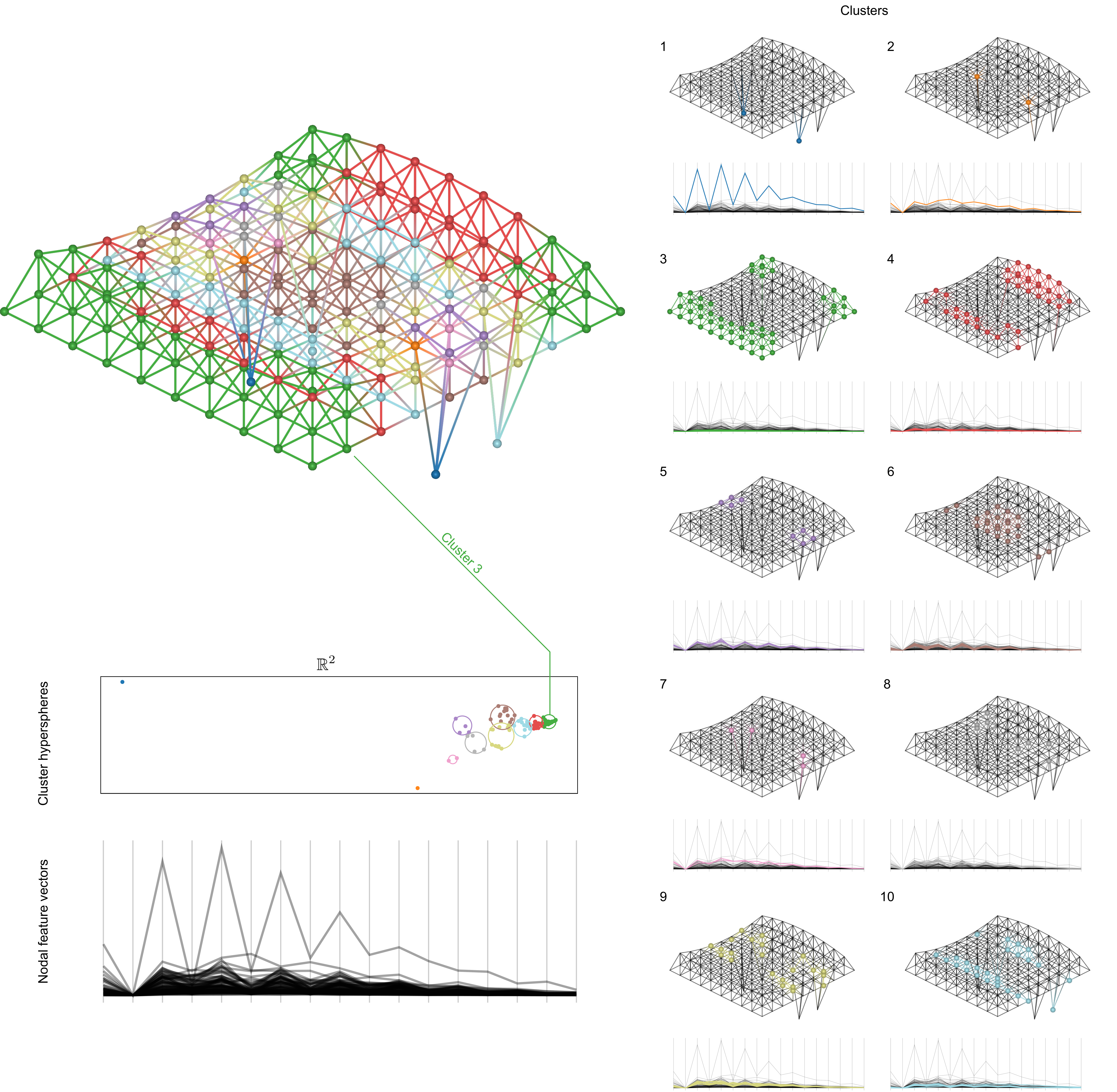}
    \caption{Clustering structural nodes into 10 similar groups. Clusters are ordered in increasing cluster complexity (hypersphere radius of clustered feature vectors). Clusters 1 and 2 contain identical nodal demands, and have a complexity score of 0.}
    \label{fig:clusters}
\end{figure}

A method of reducing the standardization penalty without extensive customization is by strategically designing fewer connections to meet the demand of multiple configurations. To achieve this, similar nodal demands must be identified to minimize the variation within each design target. We use K-means clustering on the nodal feature vectors analyzed in Section \ref{sec:3} to reduce the nodes into 10 similar groups; the results are shown in Figure \ref{fig:clusters}. The 2D embedded points are coloured by their cluster assignments along with the individual bounding spheres for each cluster. An immediate measure of the reduction in nodal complexity is the change in size of the overall bounding hypersphere to those of the individual clusters. The superimposed feature vectors, as well as highlighted feature vectors for each cluster are shown as parallel coordinate plots.

The symmetry of the design and its loading condition is recognized, with each feature vector having at least one identical feature vector across the symmetry plane between the supports. Due to the odd nodal grid spacing transverse to this plane, nodes along the centerline between the supports do not have a symmetric twin, but are clustered with their adjacent nodes as their geometries and internal forces remain similar. Symmetric node groups with highly unique force demands, such as the large compression forces experienced by the dark blue supports, are isolated in their own cluster, with an associated zero radius hypersphere due to their identical loading conditions. Groups with higher numbers indicate larger variation within the cluster, but all feature vectors show similar patterns of peaks and troughs in their highlighted parallel coordinate plots, with small variations of magnitude in each component.  

\begin{figure}[h]
    \centering
    \includegraphics[width = 0.5\textwidth]{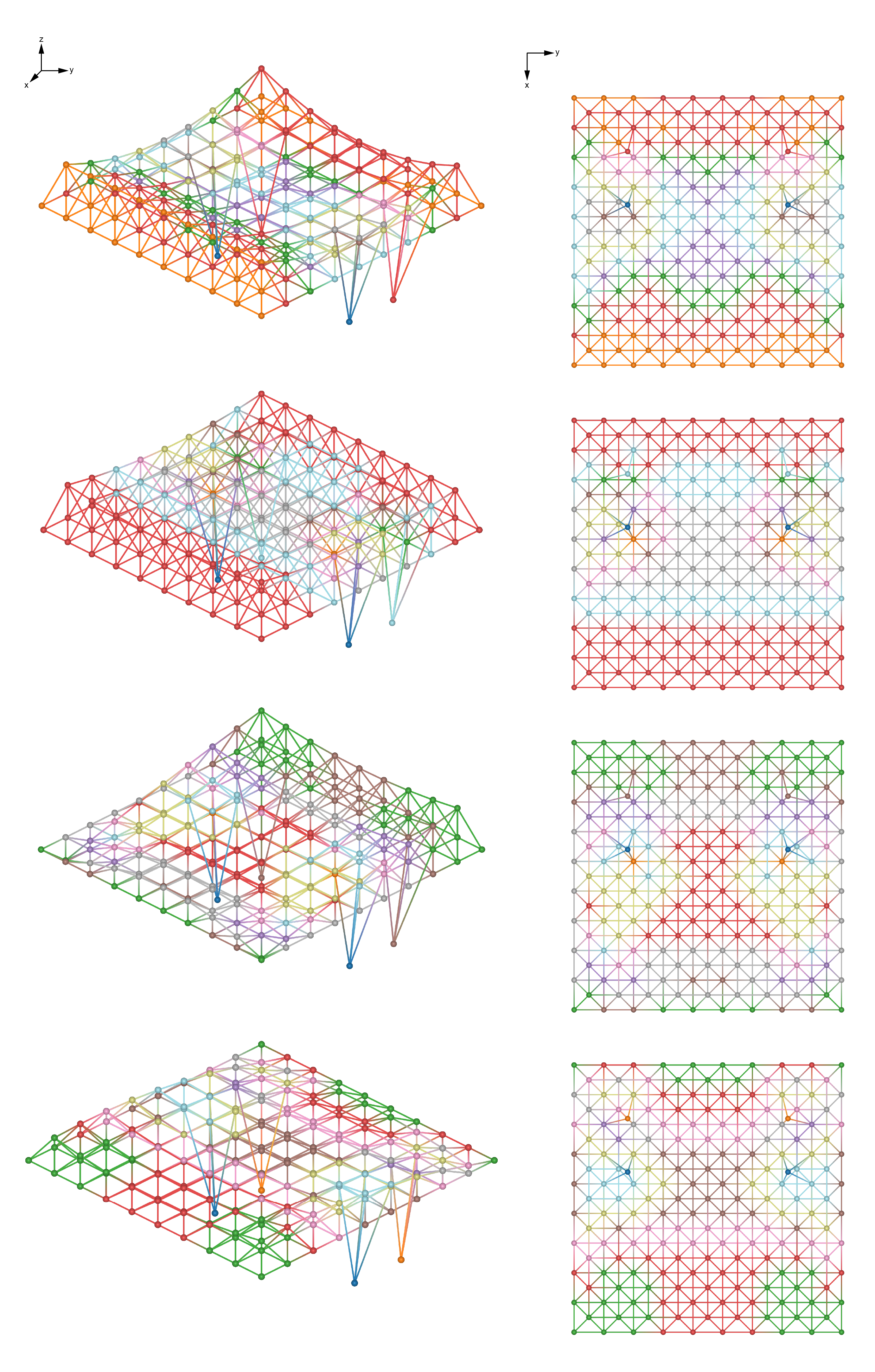}
    \caption{Four designs with 10 nodal clusters. The colours of the clusters are ordered in increasing complexity as in Figure \ref{fig:clusters}. Despite the same topology, best performing clusters are not consistently located in the structure.}
    \label{fig:4clusters}
\end{figure}

An additional four structures and their clustered node groups are shown in Figure \ref{fig:4clusters}. Clusters are coloured in the same increasing hypersphere radius order as shown in Figure \ref{fig:clusters}. Symmetry is again recognized, with the two compression supports being uniquely separated in all cases. During design iterations of the same topology, observing this consistent grouping provides an indicator that certain regions of the structure are deserving of specialized connection design. This is further evident by observing the embedded 2D feature vector points in Figure \ref{fig:clusters}, where simply providing two connection designs---one for the compression supports and one for all other nodes--would drastically reduce the overall nodal complexity score for the entire structure. It is also from the colour ordering of the clusters that despite a consistent topology and loading condition, the most similar nodes are not consistently located across the structure. For example, the green clusters shift from the central region to the structure to the cantilevered corners between two design iterations.

\begin{figure}[h]
    \centering
    \includegraphics[width = \textwidth]{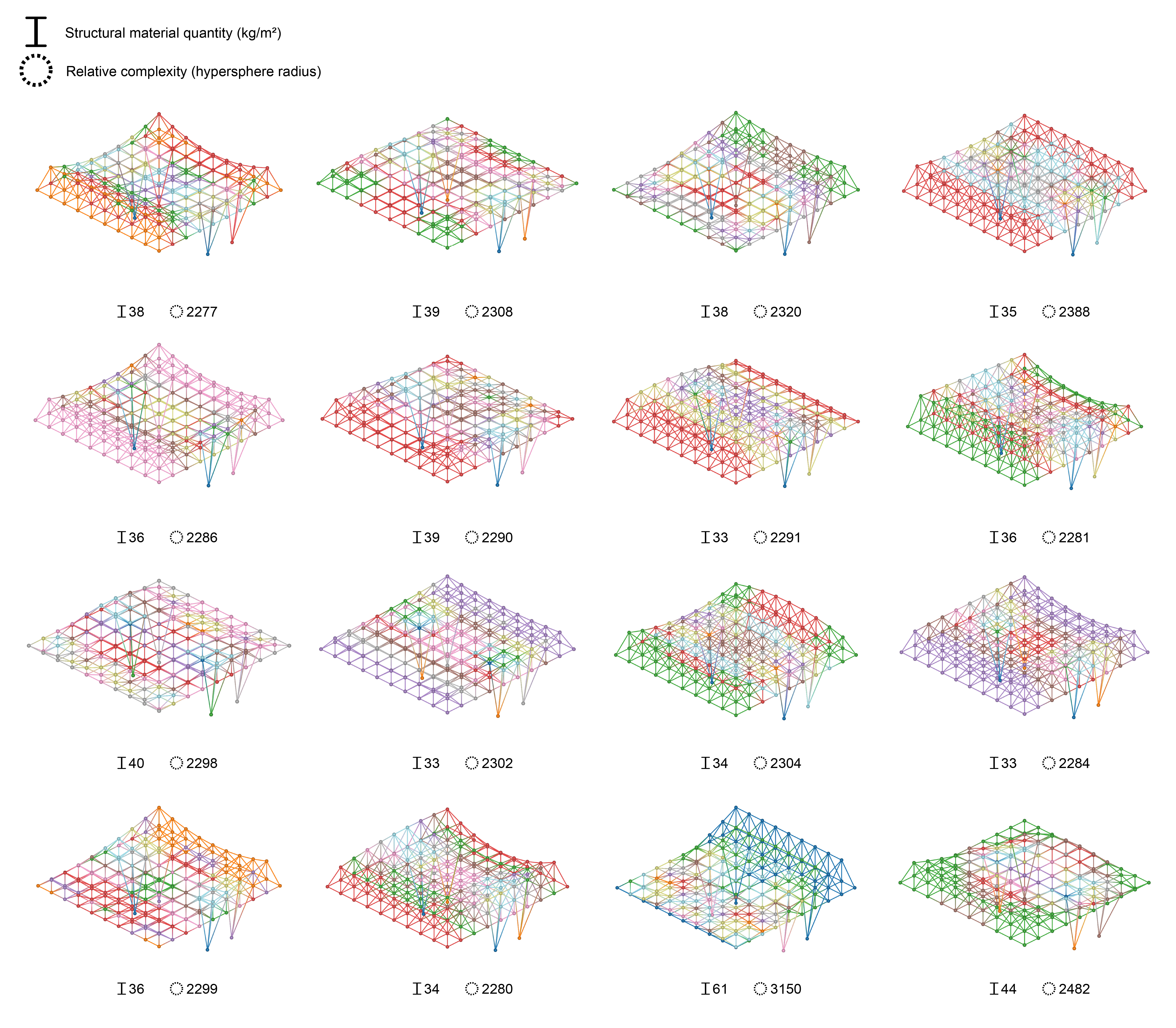}
    \caption{16 designs with a narrow range of both structural perfomance and nodal complexity. Designers are not restricted in achievable forms when considering performance metrics; rather, they can make more informed decisions when selecting a final design.}
    \label{fig:options}
\end{figure}

Despite significant variation of performance observed in Figure \ref{fig:pareto}, we note a dense region of designs with both similar complexity performance and structural performance, 16 representative samples are shown in Figure \ref{fig:options}. It is clear that even with an added performance constraint when evaluating designs, a large variation in architectural form can be achieved, as observed by the relatively small changes in both performance scores. The nodal complexity analysis can help designers narrow down and refine a wide range of designs while still maintaining design freedom. 

\begin{figure}[h]
    \centering
    \includegraphics[width = \textwidth]{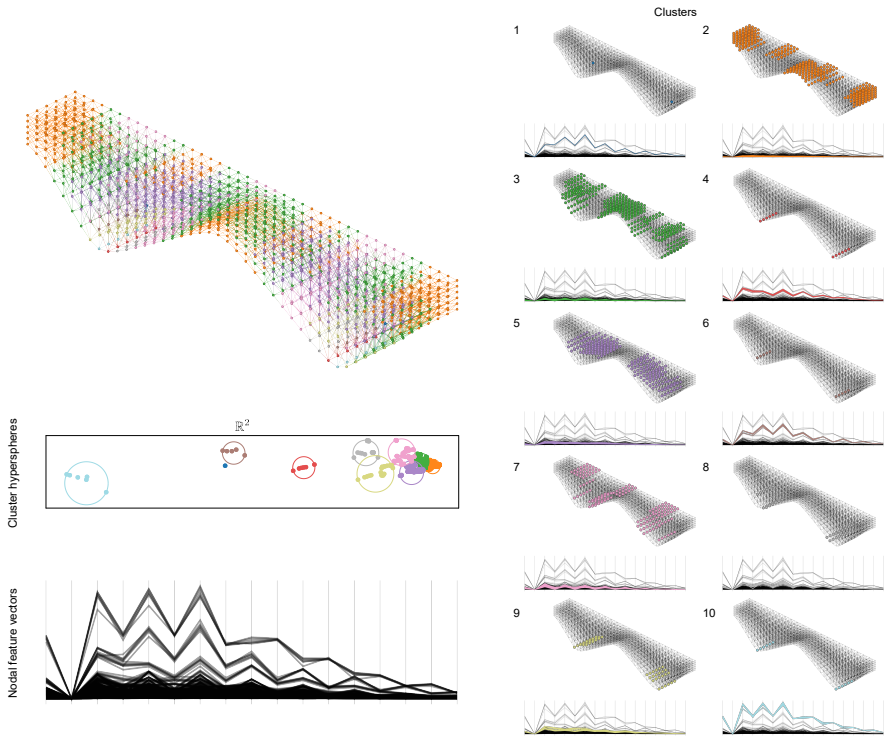}
    \caption{Clustering analysis of 3D printed lattice structure with 1548 nodes and 6427 elements. At least three distinct and suitable candidates for clustering are observed by the distinct bands in the superimposed parallel coordinate plot. They are captured by Clusters 1,4/5, and 10.}
    \label{fig:bigbridge}
\end{figure}

We perform another clustering analysis on a higher density structure with 1548 nodes and 6427 elements in Figure \ref{fig:bigbridge} (design from \cite{tam2018fabrication}). The structure was designed for 3D printed linear extrusion using PLA, with overall dimensions of $40\times 5 \times 8.5$cm. A 1N load was placed at each node for structural analysis, and supports are placed along the two lowest lines of nodes (captured by Cluster 4, red). 

We observe again one cluster of identical nodes, as well as the detection of structural symmetry. Further, we observe three natural candidates for clustering by the bunched parallel coordinate plot lines that are distinct from other feature vectors. By visual inspection of the parallel coordinate plots, it is possible to deduce the ideal starting number of clusters to minimize joint complexity while designing as few unique joints as possible. In the case of 3D printed lattice structures that lack discrete joints, the nodal demand analysis and subsequent clustering may be used in identifying critical joint regions and optimizing the extrusion path to reinforce these joints with varying amounts of material.

\subsection{Other applications}\label{subsec:4_3}
The presented method can be easily extended to multiple load cases when a more in-depth analysis is performed. Nodal feature vectors for each load case can be concatenated into feature matrices, with the same $L_2$-norm distance matrix. The feature matrix is more akin to the initial methodology by \cite{kazhdan2003rotation}, where each shape contains multiple spherical functions that capture different features of the input geometry. 

As the spherical harmonic node descriptor captures both force magnitude and geometry, it can also be used to capture the force capacity and ideal member orientations of existing structural components. \cite{amtsbergstructural} processed the forks of felled trees to act as structural nodes in a timber grid shell. The matching of tree fork geometries to the nodes of the design  required all tree forks to be of the same valence as the nodes of the structure (three branches), and did not consider the comparison of fork force capacity and the expected internal forces of the design. By representing tree fork capacities and geometries using the presented method, both the feature vectors of the tree forks can be matched to the feature vectors of the designed structural nodes irrespective of valence and with consideration of force magntiudes.

For conventional spatial trusses, the nodes of a decommissioned structure can be indexed by their representative feature vectors before storage, and can later be recalled to optimally fit the expected forces and member orientations of a new spatial truss design. An optimal assignment of an existing node to the proposed structure is one that minimizes the distance between the nodal capacity feature vector of the existing node and the nodal demand feature vector of the new design. Because only the feature vector is required to capture both force magnitudes and orientations, databases of existing structural components can be readily maintained with low memory storage requirements.

\section{Conclusion and future work}\label{sec:5}
We present an application of the spherical harmonic shape descriptor in the domain of nodal force demands in a spatial truss structure. By representing an arbitrary number of forces acting at a node as a single spherical surface, and by characterizing this surface with the spherical harmonic feature vector, we develop a consistent characterization of nodal force demands and a measure of inter-nodal dissimilarity. By representing the range of this dissimilarity by the size of the minimal bounding hypersphere in $n$ dimensional feature vector space, we provide a metric of relative nodal complexity when evaluating multiple designs. Further, we use the distance matrix of the nodal feature vectors in a given structure as the basis of node clustering, which can minimize the penalty of connection standardization and allow for the strategic design of a reduced set of customized components.

Two questions are left to be resolved, and will be the focus of future work. First, the practical materialization of a single custom connection for similar, but not identical, nodal force demands is not trivial. Additive manufacturing in conjunction with multiple load case topology optimization should be investigated as method of designing and manufacturing truss nodes with slight variations in force magnitudes and orientations. Second, as discussed in Section \ref{subsec:3_3}, the map of nodal force demands to feature vectors is not bijective, and thus computational optimization in the feature vector space is not straightforward. We intend to investigate the efficiency of numeric autodifferentiation tools to provide a similar pathway of altering a design to reduce nodal complexity as in \cite{koronaki_rationalization_2020}.

The ability to characterize structural nodes with arbitrary loading conditions while capturing both force magnitude and relative member orientations allows for greater insight during the design of spatial truss structures. As digital tools facilitate rapid design iterations, they should also provide more metrics of performance that recognize practical constraints of time, tooling, and cost. To minimize material consumption and enable a circular economy of building materials, we expect the proliferation of form and element optimized spatial truss structures with discrete nodal connections. These nodes will play a critical role in improving construction efficiency, by acting as registration devices, and in deconstruction and reuse, by reducing permanent miter-and-weld connections. We provide a method of characterizing the complexity in the demands of these discrete nodal connectors, and methods towards its strategic reduction.

\section*{Statements and Declarations}
The authors have no competing interests, financial or otherwise, pertaining to this work.

\bibliography{references}

\end{document}